\documentclass{elsarticle}
\pdfoutput=1
\usepackage{lineno,hyperref}
\modulolinenumbers[5]

\usepackage{color}

\journal{Journal of \LaTeX\ Templates}

\usepackage{array}
\usepackage{amsbsy}
\usepackage{amsmath}
\usepackage{amssymb}
\usepackage{makeidx}
\usepackage{graphicx}
\usepackage{listings}
\usepackage{tikz}
\usetikzlibrary{arrows.meta}
\usetikzlibrary{shapes,arrows}

\usepackage{amsmath}
\usepackage{amssymb}
\usepackage{makeidx}
\usepackage{graphicx}
\usepackage{color}

\newcommand{\wrt}{w.r.t.\ }

\newcommand{\rhs}{rhs.\ }

\newcommand{\hide}[1]{}

\newcommand{\up}[1]{ ^{(#1)}}
\newcommand{\inv}[1]{ ^{-#1}}

\newcommand{\order}[1]{\mathcal{O}(#1)}

\newcommand{\nm}{{\rm\,nm}}

\newcommand{\myvec}[1]{\vec{#1}}
\newcommand{\vr}{{\myvec{r}}}

\newcommand{\vc}{{\myvec{c}}}

\newcommand{\vk}{{\myvec{k}}}

\newcommand{\vA}{{\myvec{A}}}

\newcommand{\ad}{a^\dagger}

\newcommand{\mymat}[1]{{\widehat{#1}}}

\newcommand{\mh}{\mymat{h}}

\newcommand{\mC}{\mymat{C}}

\newcommand{\mH}{\mymat{H}}

\newcommand{\mL}{\mymat{L}}

\newcommand{\mM}{\mymat{M}}

\newcommand{\md}{\mymat{d}}
\newcommand{\mO}{\mymat{O}}
\newcommand{\mP}{\mymat{P}}

\newcommand{\mR}{\mymat{R}}
\newcommand{\ms}{\mymat{s}}
\newcommand{\mS}{\mymat{S}}

\newcommand{\mU}{\mymat{U}}
\newcommand{\mV}{\mymat{V}}

\newcommand{\cA}{\mathcal{A}} 
\newcommand{\cH}{\mathcal{H}} 
\newcommand{\cI}{\mathcal{I}} %
\newcommand{\cN}{\mathcal{N}} %
\newcommand{\RR}{\mathbb{R}}
 
\newcommand{\one}{\mathbf{1}} 

\newcommand{\adj}{^\dagger}

\renewcommand{\r}{\rangle}
\renewcommand{\l}{\langle}

\newcommand{\sd}{\downarrow}

\newcommand{\ddt}{\frac{d}{dt}}

\newcommand{\om}{\omega}
\newcommand{\si}{\sigma}

\newcommand{\ep}{\epsilon}
\newcommand{\al}{\alpha}

\newcommand{\be}{\beta}

\newcommand{\ga}{\gamma}
\newcommand{\de}{\delta}

\renewcommand{\th}{\theta}

\newcommand{\lcase}{\left\{\begin{array}{ll}}
\newcommand{\rcase}{\end{array}\right.}
\renewcommand{\bar}{\begin{array}{ll}}
\newcommand{\ear}{\end{array}}
\newcommand{\bal}{\begin{align}}
\newcommand{\eal}{\end{align}}
\newcommand{\bma}{\begin{pmatrix}}
\newcommand{\ema}{\end{pmatrix}}
\newcommand{\beq}{\begin{equation}}
\newcommand{\eeq}{\end{equation}}
\newcommand{\bel}[1]{\begin{equation}\label{eq:#1}}
\newcommand{\eel}{\end{equation}}
\newcommand{\bea}{\begin{eqnarray}}
\newcommand{\eea}{\end{eqnarray}}
\newcommand{\beaNN}{\begin{eqnarray*}}
\newcommand{\eeaNN}{\end{eqnarray*}}

\newcommand{\Ef}{\mathcal{E}}
\newcommand{\vEf}{\vec{\mathcal{E}}}

\newcommand{\q}{\!/\!}

\newcommand{\alp}{{\al_\perp}}
\newcommand{\bep}{{\be_\perp}}

\newcommand{\lcode}[1]{{\tt\detokenize{#1}}}

\newcommand{\revised}[2]{{} {#2}}

\begin{document}

\begin{frontmatter}



%
%

\title{The hybrid anti-symmetrized coupled channels method (haCC) for the tRecX code}
 
\author{Hareesh Chundayil$^1$, Vinay P. Majety$^1$\footnote{\textit{E-mail address:} vinay.majety@iittp.ac.in}
, and Armin Scrinzi$^2$\corref{correspondingAuthor}}
\address{$^1$ Department of Physics and CAMOST, Indian Institute of Technology Tirupati, Yerpedu, India}
\address{$^2$ Ludwig Maximilian University, Theresienstrasse 37, 80333 Munich, Germany}
\cortext[correspondingAuthor] {\textit{E-mail address:} Armin.Scrinzi@lmu.de}

\begin{abstract}
We present a new implementation of the hybrid antisymmetrized Coupled Channels (haCC) method in the 
framework of the tRecX [A. Scrinzi, Comp. Phys. Comm.,  270:108146, 2022.]. 
The method represents atomic and molecular multi-electron functions by combining CI functions,  
Gaussian molecular orbitals, and a numerical single-electron basis. It is suitable for describing high harmonic generation and the strong-field dynamics of ionization. Fully differential photo\-emission spectra are computed by the tSurff method. 
The theoretical background of haCC is outlined and key improvements compared to its original formulation are highlighted. We discuss control of over-completeness resulting from the joint use of the numerical basis and Gaussian molecular orbitals 
by pseudo-inverses based on the Woodbury formula. Further new features of this tRecX release are the iSurff method, new input features, 
and the AMOS gateway interface. The mapping of haCC into the tRecX framework for solving the time-dependent Schr\"odinger equation is shown.   
Use,  performance, and accuracy of haCC are discussed on the examples of high-harmonic generation and strong-field photo-emission by short
laser pulses impinging on the Helium atom and on the linear molecules $N_2$ and $CO$.
\end{abstract}

\begin{keyword}
\texttt Schr\"odinger solver  \sep strong field physics \sep attosecond physics \sep recursive structure
\end{keyword}

\end{frontmatter}
\begin{small}
\noindent
{\bf PROGRAM SUMMARY}
\\
{\em Program title:} tRecX --- time-dependent Recursive indeXing (tRecX=tSurff+irECS)
\\
{\em CPC Library link to program files:} (to be added by Technical Editor)
\\
{\em Developer's repository link:} https://gitlab.physik.uni-muenchen.de/AG-Scrinzi/tRecX 
\\
{\em Licensing provisions:} GNU General Public License 2
\\
{\em Programming language:} C++
\\
\revised{}{{\em Libraries:} Eigen, arpack, lapack, blas, boost, FFTW (optional)} 
\\
{\em Journal Reference of previous version:} A. Scrinzi, Comp. Phys. Comm.,  270:108146, 2022.
\\
{\em Does the new version supersede the previous version: } Yes
\\
{\em Reasons for the new version:} Major new functionality: haCC --- hybrid antisymmetrized coupled channels method
\\
{\em Summary of revisions:} Main additions are haCC and iSurff. Code usage and compilation were improved.
\\
{\em Nature of problem}: tRecX is a general solver for time-dependent Schr\"odinger-like problems, 
with applications mostly in strong field and attosecond physics.  There are no technical restrictions on 
the spatial dimension of the problem with up to 6 spatial dimensions realized in the strong-field double 
ionization of Helium.  Gaussian-based quantum chemical multi-electron atomic and molecular structure can be
combined with the numerical basis. A selection of coordinate systems is available and any Hamiltonian 
involving up to second derivatives 
and arbitrary up to three dimensional potentials can be defined on input by simple scripts.
\\
{\em Solution method:} The method of lines is used with spatial discretization by a flexible combination of one dimensional basis sets, 
DVR representations, discrete vectors, expansions into higher-dimensional eigenfunctions of user-defined operators, 
and Gaussian based molecular orbitals.
Multi-electron Gaussian-based CI (configuration interaction) functions for neutrals and ions are combined with the numerical basis.
Photo-emission spectra are calculated using the time-dependent surface flux method (tSurff) in combination with infinite 
range exterior complex scaling (irECS) for absorption. The code is object oriented and makes extensive use of tree-structures and recursive algorithms. 
Parallelization is by MPI. Code design and performance allow use in production as well as for graduate level training.
\end{small}

\tableofcontents

\section{Introduction}

The ``hybrid anti-symmetrized coupled channels'' (haCC) method unites  quantum chemical multi-electron structure with a purely numerical description of non-perturbative strong field interactions. In its first formulation it was presented in \cite{majety15:hacc} and it was used for applications
that showed the essential role of anti-symmetrization and correlation for strong-field ionization rates \cite{majety15:dynamicexchange}, for fully differential 
photoemission from linear molecules \cite{majety17:co2spectra}, 
for benchmark ionization rates of atoms and small diatomics \cite{majety15:static}, and for demonstrating
the absence of correlation effects in the attosecond delays in the photo-emission by elliptically polarized light \cite{majety17:heElliptic}.
The present major upgrade of public domain tRecX code \cite{scrinzi22-trecx} includes an improved version of haCC with all capabilities 
cited above. 

The tRecX package is designed to be a high-performance, yet flexible and robust code with good maintainability and usability
for Schr\"odinger-like time-dependent problems, where the original design was for applications in strong-field and attosecond physics. 
Its main use has been for computing the interaction of atomic and molecular systems
in non-perturbatively strong laser fields. It implements a range of techniques such as irECS (infinite-range exterior complex
scaling \cite{Scrinzi2010}), tSurff (the time-dependent surface flux method \cite{Tao2012,scrinzi12:tsurff}), general and mixed
gauges \cite{majety15:gauge}, and the FE-DVR method for complex scaling \cite{rescigno00:fem-dvr,weinmuller17:dvr}. 
The most demanding applications of tRecX have been the computation of fully-differential 
double electron emission spectra of the Helium atom \cite{zielinski16:doubleionization,zhu20:doubleionization} at laser wave length 
from 10 to 800 nm. These capabilities were included with the  first official release of tRecX.
The applications for multi-electron systems \cite{majety15:dynamicexchange,majety17:co2spectra,majety15:static,majety17:heElliptic} 
with arbitrary alignment between the direction of laser polarization and the molecular axis have not been part of the public code so far.

The new release keeps with the original tRecX design.  
A conscious effort of adhering to good programming practice is being made for ensuring re-usability and maintainability. The C++ code
uses abstract and template classes for uniform and transparent code structure. The classes reflect concepts that are familiar in physics 
such as the linear and more specifically Hilbert space, operators that are usually but not necessarily linear maps, and wave functions.
Discretization of the wave function is in terms of an abstract basis set class, whose specific implementation
covers the whole range from discrete sets of vectors, over grids, finite-elements, standard basis sets 
such as spherical harmonics, all the way to expansions in terms of eigenfunctions of a user-defined operator. With the present release, Gaussian-based molecular orbitals and multi-electron CI functions were added to that list. 
These are combined in a tree-structured hierarchy that admits building correlated (non-product) bases from one- and multi-dimensional factors.
For performance, numerical libraries such as Lapack \cite{lapack}, Eigen \cite{eigen}, and FFTW \cite{fftw} are used on the low level. Parallelization
is through MPI with automatic load-balancing according to  self-measurement of the code. 
Also, non-trivial model Hamiltonians can be implemented quickly with little compromise in computational performance. Details of the tRecX structure and algorithms can be found in Ref.~\cite{scrinzi22-trecx}.

The present paper introduces the new formulation of haCC, explains key elements of its efficient implementation and how it fits into the
wider tRecX framework. We show how haCC is invoked from tRecX, and how one can assess convergence and accuracy of a haCC calculation. Further, the new addition of the iSurff extension \cite{morales16:isurf} to the tSurff method is explained and demonstrated by examples. Various
improvements in usability are highlighted as appropriate.

\section{The haCC method}
\label{sec:haCC}

The purpose of haCC is to solve the Schr\"odinger equation for a multi-centered multi-electron system --- a molecule --- in presence of a dipole electric field with general polarization of the field vector $\vEf(t)$. The Schr\"odinger equation for the problem has the form,  in length gauge, 
\beq
i \ddt \Psi(t) = \left[H_0+\vEf(t)\cdot\sum_{n=1}^N\vr_n\right] \Psi(t),
\eeq
where $H_0$ is the Hamiltonian of the field-free $N$-electron system. We use atomic units (au), where electron mass, Planck constant, and elementary charge are all =1: $m_e=\hbar=e^2=1$

When $\vEf(t)$ becomes large, at low laser frequencies $\om$, or on very short time scales there is typically a broad spectrum of continuum states involved in the dynamics that defies common expansion into standard basis functions. For systems with up to two electrons, expansion on grids or localized basis functions such as B-splines, finite elements (FE), or FE-DVR is feasible. For systems with more electrons and when the majority of electrons remains in near bound states,  one needs to supplement the description by methods that are better suitable for near-bound electrons. Examples for this type of approach are the time-dependent R-matrix \cite{RMT2020}, B-spline R-matrix \cite{zatsarinny2005}, and the XChem \cite{XCHEM2023} codes.  

\subsection{The haCC expansion}

The central assumption of haCC is that the dynamics is dominated by the motion of a single electron (called the ``free electron'' in the following) 
and that the correlation of this motion with the remaining electrons can be described by
coupling between the first few excited ionic channels. In addition to that, the fully correlated field-free neutral ground state and possibly neutral excited states are included. This leads to an ansatz for the 
wave function in the form
\beq\label{eq:hacc}
|\Psi\r = \sum_{\cN} |\cN\r c_\cN + \sum_{A}\left\{\sum_i \ad_i|A\r c^A_i + \sum_\al \ad_\al|A\r c^A_\al\right\}.
\eeq
\revised{}{
The field-free neutral $|\cN\r$ states and the ionic $|A\r$ states are CI (configuration interaction) functions. Both sets of states are constructed from the same $I$ Gaussian-based molecular orbitals $|i\r, i=0,\ldots I-1$. The free electron is described by an expansion into the molecular orbitals $|i\r$ and numerical basis functions $|\alp\r$ with the creation operators $\ad_i$ and $\ad_\al$, respectively. The $|\alp\r$ are
explicitly orthogonalized to the $|i\r$ as}
\beq\label{eq:alfaBasis}
|\alp\r=\left( 1-\sum_{i=0}^{I-1}|i\r\l i|\right)|\al\r, \qquad |\al\r = |Y_{m_\al l_\al}\r |\xi^{i_\al}_{n_\al}\r.
\eeq
The $ \l \phi,\th|Y_{m l}\r=Y_{ml}(\phi,\th)$ are standard spherical harmonics. For the radial discretization we use a FE-DVR method \cite{rescigno00:fem-dvr,weinmuller17:dvr}, where the  $\l r|\xi^i_n\r=\xi^i_n(r)$ are Lagrange polynomials at the Lobatto quadrature points $r_i$ for the radial interval $r\in[r_n,r_{n+1}]$. On the last,  semi-infinite interval $|r_N,\infty)$ we use Lagrange-polynomials  times an exponential factor $\exp(-\kappa r)$ \cite{Scrinzi2010} and a Radau quadrature that include 
\revised{$R_N$}{$r_N$} as one of the quadrature points. The quadratures are used for the numerical calculation of integrals.
By expressing multi-electron functions through the creation operators we ensure anti-symmetrization of the free-electron basis with the ionic states $|A\r$. Neutral and ionic functions can be, in principle, drawn from any Gaussian-based quantum chemistry code. As discussed below, the computation of matrix elements requires rather detailed access to reaction density matrices and generalized Dyson orbitals, which at present we obtain through a customized version of the COLUMBUS code \cite{lischka11:columbus}.

An important feature of the $\al$-basis is that the interval boundaries $r_n$ can be freely chosen and the size of the angular expansion and degree of the Lagrange polynomials can be adjusted 
to \revised{the needs of the specific system on each individual interval.}{the local properties of the wave function in a specific system.}
\revised{ In particular, the molecular orbitals $|i\r$ will have cusp-like behavior at radii where an atom is located. In the vicinity 
of these radii small intervals and high expansion orders will be chosen for the $|\al\r$. 
This allows to appropriately render the free electron's scattering from the atom.}
{
In particular, at the location of the atoms, both, bound and scattering parts of the solution will have cusp-like peaks. While the bound state part of the peaks is largely captured by the Gaussians of the molecular orbitals $|i\r$, peaks in the scattering part must be representable by the numerical functions
$|\al\r$. By using small shells $[r_n,r_{n+1}]$ around the radial position of the atoms and a large angular expansions on those intervals, one can represent the peaks well without inflating the basis at larger distances from the atoms. 
}

A very similar ansatz is used in Refs.~\cite{XCHEM2023,XCHEM2023pub}.
The important distinction of the ansatz (\ref{eq:hacc}) from that expansion is our use of the dense numerical $\al$-basis in the complete space including the domain of the molecular orbitals $|i\r$. This choice was made for being able to describe non-perturbative strong-field interactions where the so-called ``recollision'' processes play a key role. 
Recollsion determines high harmonic generation and photo-electron emission by long wavelength laser fields:
the field drives electrons into large spatial \lcode{excursions of 10 au and more}{of several 10s of au} , after which they recollide  with their parent ion
at a broad range of energies. The $\al$-basis provides the necessary flexibility for describing this dynamics. It comes at the cost of computing two-electron integrals involving both, Gaussian-based molecular  orbitals $|i\r$ and the numerical functions $|\al\r$, see Sec.~\ref{sec:computation}.

\subsection{Gauge}

The expansion functions of haCC are chosen with the physical model assumption that ground state and
first few excited states of neutral and ion play an essential role also when the external field $\vEf(t)$ is present. 
Upon changing from length to velocity gauge all states, including the states considered essential, 
become time-dependent by the transformation
factor $\exp[-i\vA(t)\cdot\vr]$ with the vector potential $\vA=\int^t_{-\infty} d\tau \vEf(\tau)$. Our assumption singles out 
length gauge as the gauge where the essential states are time-independent. 
This difference is particularly important in strong fields or at low laser frequencies $\om$, 
where the phases $\vA\cdot\vr\approx\vEf\cdot\vr/\om$ vary significantly across the extension of the essential states. \

For avoiding the costly re-computation of matrix elements at each time $t$, we use length gauge up to the radius where molecular orbitals become negligible. For reasons of numerical efficiency, we switch to a mixed gauge beyond that radius that asymptotically coincides with the velocity gauge. That gauge is constructed such that field-dependence of the matrix has the form of multiplications by $\Ef_i(t)$, $\vA_i(t)$, and $A_i(t)A_j(t)$ with $i,j\in\{x,y,z\}$. This obviates the need for re-computing
matrix elements at each step at the expense of additional quadrupole-like terms. With $z$-polarization, the interaction operator
has three parts that are multiplied by  $\Ef_z(t)$, $A_z(t)$, and $A^2_z(t)$ respectively, in general polarization it consists of 13 terms.  Details of the approach and demonstration of its essential role for computing strong field effects can be found in \cite{majety15:gauge}.

\subsection{Representation of operators}
\label{sec:operatorImplementation}

Our discretization space can be written as the direct sum  $\cH=\cN\oplus\cI\oplus\cA$ 
of the neutral CI space $\cN$ and the spaces where the ionic functions $|A\r$ are augmented by molecular orbitals $\cI=\text{span}(\ad_i |A\r)$ and
by the orthogonalized numerical $\al$-functions $\cA=\text{span}( \ad_\al |A\r)$.  Any operator matrix $\mM$, including the overlap matrix, can be written in block-matrix form
\beq
\mM=\bma
\mM_{\cN\cN} &\mM_{\cN\cI} &\mM_{\cN\cA} \\
\mM_{\cI\cN} &\mM_{\cI\cI} &\mM_{\cI\cA} \\
\mM_{\cA\cN} &\mM_{\cA\cI} &\mM_{\cA\cA} 
\ema.
\eeq
As written, each block is a full matrix except for some sparsity in the overlap matrix and possible zeros due to global symmetries like angular momentum. The exchange term of the Coulomb interaction will always create a full matrix. However, because of the orthogonalization of the $|\alp\r$
to the globally defined orbitals $|i\r$ also the blocks $\mM_{\cA\cA}$ of any single particle operator and even the overlap block 
$\mO_{\cA\cA}$ are full. Given the large dimension of, say, $\dim(\cA)\sim 10^5$ and our need to separately apply multiple time-dependent operators of the mixed gauge dipole interaction this is a severe computational disadvantage. 
The fill in of the matrices originates from comparatively few molecular orbitals $\lesssim50$, and it can therefore
replaced by applying sparse matrices in combination with operators of rank $\lesssim50$, which is computationally efficient.

\newcommand{\mm}{\widehat{m}}
\newcommand{\mo}{\widehat{o}}
One finds that matrix elements $M_{\al j}$ \wrt the orthonormal basis $|\alp\r$ can be obtained from matrix elements 
$m\up{1}_{\al l}$ for the local basis functions $|\al\r$ without the orthonormalization by 
\beq
M_{\al j} = m\up{1}_{\al j} + \sum_k U_{\al k}m\up{1}_{kj}\qquad\text{with}\quad U_{\al k}:=-\l \al|k\r
\eeq
and similarly
\beq\label{eq:operatorHaCC}
M_{\al \be} = m\up0_{\al \be} 
+ \sum_k U_{\al k}m\up{2}_{k\be}
+ \sum_k m\up2_{\al k}U_{k\be}
+ \sum_{kl} U_{\al k}m\up2_{kl}U_{l\be}.
\eeq
The form of the $\mm\up{x}$ will be discussed below.

In block-matrix form the \revised{}{part of the operator pertaining to the ionic channels $\cI\oplus\cA$} reads
\beq
\bma\label{eq:operatorHaCCblock}
\mM_{\cI\cI} & \mM_{\cI\cA}\\ 
\mM_{\cA\cI} & \mM_{\cA\cA} 
\ema
= \mU \mm \mU\adj
\eeq
with
\beq\label{eq:UMU}
\mU=\bma
1 & 0 & 0\\
0 & 1 & \mU_{\cA\cI}
\ema
\qquad\text{and}\qquad
\mm=\bma
\mm\up0_{\cI\cI}&\mm\up1_{\cI\cA}&\mm\up1_{\cI\cI}\\
\mm\up1_{\cA\cI}&\mm\up0_{\cA\cA}&\mm\up2_{\cI\cA}\\
\mm\up1_{\cI\cI}&\mm\up2_{\cA\cI}&\mm\up2_{\cI\cI}
\ema
\eeq
The transformations $\mU$ involve only matrices where one dimension is equal to the number of orbitals \revised{}{$\dim(\cI)=I\lesssim50$}
and $\mm\up0_{\cA\cA}$ is sparse except for the exchange term. 
The \lcode{class OperatorHaCC} that implements this type of operators is discussed below.

\subsection{Inverse overlap and over-completeness}
\label{sec:overcompleteness}
\revised{The time-evolution of the expansion coefficients $\vc(t)$ is given by}
{
Denoting the expansion coefficients of Eq.~\ref{eq:hacc} by the vector $\vc=(c_\cN,c_i^A,c_\al^A)$ and the time-dependent Hamiltonian matrix by $\mH(t)$, 
 the time-evolution of the $\vc$ is given by
}
\beq
\ddt \vc(t) = -i \mO\inv1 \mH(t) \vc(t).
\eeq
In haCC, the overlap matrix $\mO$ has the structure
\beq
\mO=\bma 
\one_\cN & \mO_{\cN\cI} & 0 \\
\mO\adj_{\cN\cI} & \one_\cI & 0 \\
0 & 0 & \mO_{\cA\cA} 
\ema
\eeq
with the identity matrices $\one_\cN$ and $\one_\cI$ on the respective subspaces. 
It is easy to see that 
\beq
\mO_{\al A,\be B}=\l A | a_\al \ad_\be| B\r=\de_{AB} \l \alp | \bep\r=\de_{AB} \l \al | \be\r- \sum_k U_{\al k} \overline{U}_{k \be}
\eeq
or in block matrix notation
\beq
\mO_{\cA\cA}=\mo_{\cA\cA}-\mU_{\cA\cI}\mU\adj_{\cA\cI}.
\eeq
The two overlap blocks on $\cN\oplus\cI$ and $\cA$ are independent. The structure of each block allows for the use of the 
Woodbury formula \cite{woodbury50} by which one writes the inverse of a matrix of the form
\beq
\mO = \mo - \mU \mU\adj
\eeq
as
\beq\label{eq:woodbury}
\mO\inv1 = \mo\inv1 - \mo\inv1 \mU \mC\inv1  \mU\adj \mo\inv1
\eeq
with 
\beq
 \mC = \mU\adj\mo\inv1 \mU - \one
\eeq
In our case, the $\mo$ for the $\cN\oplus\cI$ and $\cA$ blocks are identity and diagonal, respectively, and the dimensions of $\mC$ are small. 
Application of the \rhs of Eq.~\ref{eq:woodbury} is thus computationally efficient. 

The overlap blocks for both, $\cN\oplus\cI$ and $\cA$, can become ill-conditioned, when 
the bases are sufficiently large. The ill-conditioning of $\mO$ re-appears as ill-conditioning of 
$\mC$. As $\mC$ is small, near-zero eigenvalues can be removed by diagonalizing and forming a   
pseudo-inverse
\beq
\mC\inv1_\ep=\mV_\ep \Sigma\inv1_\ep \mV\adj_\ep,
\eeq
where eigenvalues $<\ep$ and the corresponding eigenvectors were removed from $\Sigma_\ep$ and $\mV_\ep$. 
The pseudo-inverse of the full matrix $\mO$
\beq\label{eq:pseudoInverse}
\mO_\ep\inv1 =  \mo\inv1 - \mo\inv1 \mU \mC_\ep\inv1  \mU\adj \mo\inv1
\eeq
now acts only on the non-singular subspace and the operator
\beq\label{eq:projEps}
\mP_\ep:=\mO \mO_\ep\inv1
\eeq
is a projector onto the subspace where $\mO$ is non-singular.

\subsection{Matrix elements}
\label{sec:computation}

Matrix elements between the various parts of the basis are best derived in second quantized notation writing single- and two-particle operators as, respectively,
\beq
S=S_{xy} \ad_x a_y
\qquad\text{and} \qquad
T=T_{xr,ys}  \ad_x  \ad_r a_y a_s, \qquad x,y,r,s \in \{ i, \al\}.
\eeq
Here and in the following we imply summation over equal index pairs.
 
The matrix elements need to be derived separately for all combinations of our subspaces $\cN,\cI$ and $\cA$.
In the derivation one can take advantage of the orthonormalization $\l j | \alp\r=0$. Only in the final step one expresses the matrix elements in terms
of the $|\al\r$ and $|i\r$.
For example, one obtains for the matrix elements of a single-particle operator  $\mS_{\cA\cA}$ on $\cA$ 
\beq
\l A| a_\al S \ad_\be |B\r = \de_{AB} S_{\al\be} + \de_{\al\be} \rho^{AB}_{ij}S_{ij},
\eeq
with the single particle reaction density matrices $\rho^{AB}_{ij}:=\l A |a_i \ad_j| B\r$.
One resolves the orthogonalization as
\beq
S_{\al\be}=\l \alp|S|\bep\r=\l \al|S|\be\r + U_{\al k} \l k | S | \be\r + \l \al | S | k \r U\adj_{\be k} +   U_{\al l}\l l | S | k \r U\adj_{\be k}
\eeq
One can read off the matrix blocks $\ms\up{0}$ and $\ms\up{2}$ entering Eq.~(\ref{eq:UMU}) in the case of single-particle operators $M=S$
\bea
\ms\up0_{\al A,\be B}&=&\delta_{AB}\l \al | S |\be\r + \delta_{\al\be} \rho^{AB}_{ij}\l i|S|j\r,
\\
\ms\up2_{u A,v B}&=&\delta_{AB}\l u | S |v\r, \quad u=\al, i,\quad v=\be,j.
\eea

An important practical complication for the haCC expansion arises from the combination  of ionic CI states with single-electron
orbitals:  for two-particle operators $T= T_{ik,jn}\ad_i \ad_k a_j a_n$ one needs the three-particle reaction density matrices 
of the ionic states $A,B$:
\beq
\l A |a_u  \ad_i \ad_k a_j a_n  \ad_v | B\r = \l A| \ad_i \ad_k \ad_v  a_u a_j a_n | B\r+\ldots
\eeq
We are using a customized version of COLUMBUS \cite{lischka11:columbus}, where we can directly access the full CI function 
to construct and save these matrices. \revised{}{The customized version of COLUMBUS is not publically available, but} data for selected systems is supplied on a repository \cite{chemical-git}. \revised{}{Data for other systems can be made available upon request.}

\subsection{Representation in tRecX}
\label{sec:representation}

The haCC expansion consists of a hierarchy of a ``hybrid'' discretizations within the recursive scheme of tRecX (see Ref.~\cite{scrinzi22-trecx}).  
Hybrid here refers to the fact that the same domains in space are discretized with different types of expansion functions. The neutral CI states $|\cN\r$ 
are defined on the same domain of multi-electron coordinates as the channel functions $\ad_i|A\r$, the orbitals $|i\r$ and the numerical functions $|\al\r$ are both defined on $\RR^3$. 

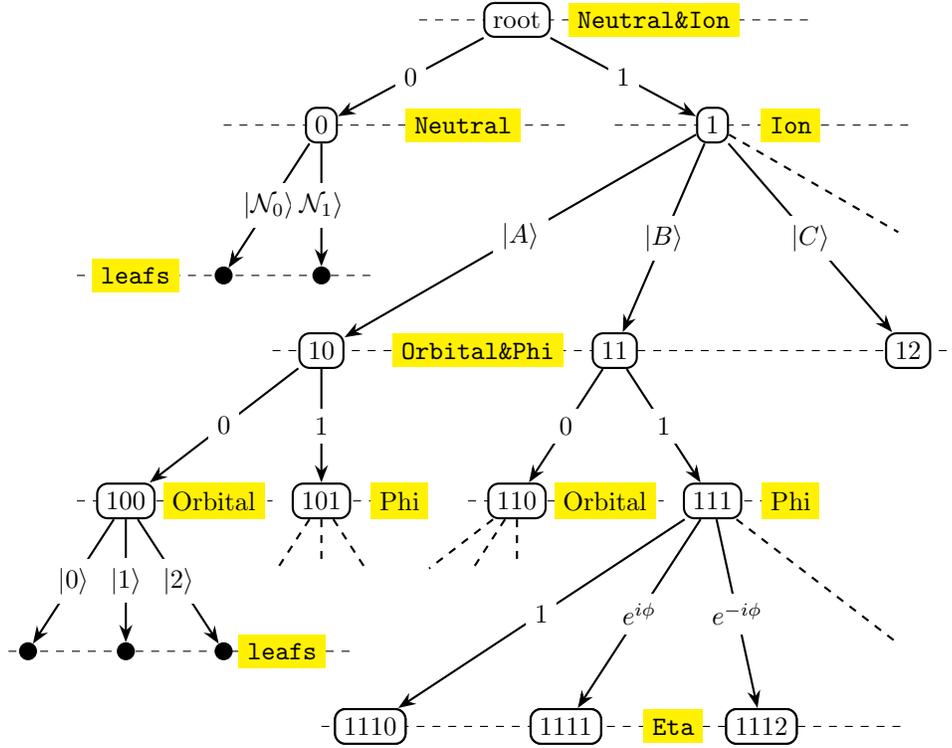
\begin{figure}
\begin{tikzpicture}
\begin{scope}
\def\n{1.3}
\def\m{-2}
\def\lni{0*\m}
\def\ln{\lni+0.7*\m}
\def\lop{\ln+1.5*\m}
\def\lo{\lop+\m}
\def\le{\lo+1.5*\m}
\def\lla{\ln+1*\m}
\def\llb{\lo+1*\m}
\def\llc{\llb+1*\m}
\begin{scope}[anchor=center,every node/.style={rectangle,fill=yellow}]

\draw[dashed] (2*\n,\lni) -- (6*\n,\lni) node[pos=0.6]{\lcode{Neutral&Ion}};

\draw[dashed] (0*\n,\ln) -- (3.5*\n,\ln) node[pos=0.7]{\lcode{Neutral}};
\draw[dashed] (4*\n,\ln) -- (7*\n,\ln) node[pos=0.6]{\lcode{Ion}};

\draw[dashed] (-1.5*\n,\lla) -- (1.5*\n,\lla) node[pos=0.2]{\lcode{leafs}};
\draw[dashed] (0.5*\n,\lop) -- (7.5*\n,\lop) node[pos=0.3]{\lcode{Orbital&Phi}};

\foreach \x in {0,4}
{
    \draw[dashed] (-1.5*\n+\x*\n,\lo) -- (0.5*\n+\x*\n,\lo) node[pos=0.7]{Orbital};
    \draw[dashed] (1*\n+\x*\n,\lo) -- (2*\n+\x*\n,\lo) node[pos=0.8]{Phi};
}
\draw[dashed] (-2.2*\n,\llb) -- (1.3*\n,\llb) node[pos=0.8]{\lcode{leafs}};

    \draw[dashed] (1*\n,\le) -- (7*\n,\le) node[pos=0.6]{\lcode{Eta}};

\end{scope}
\begin{scope}[every node/.style={rectangle,rounded corners, thick,draw, fill=white}]
\tikzstyle{leaf} = [circle, minimum width=6pt, fill=black, inner sep=0pt]
\tikzstyle{dots} = [draw=none]

    \node (0) at (3*\n,0) {root};
    \node[leaf] (000) at (0*\n,\lla) {};
    \node[leaf] (001) at (1*\n,\lla) {};
    
    \foreach \c in {0,1}
        \node (0\c) at (1*\n+\c*4*\n,\ln) {\c};
    
    \foreach \p in {0,1,2}
        \node (01\p) at (1*\n+\p*3*\n,\lop) {1\p};
        
    \foreach \l in {0,1}
    {
        
        \foreach \q in {0,1}
            \node (01\l\q) at (-\n+\l*4*\n+\q*2*\n,\lo) {1\l\q};

        \foreach \k in {0,...,2}
            \node (0111\k) at (1.5*\n+\k*2*\n,\le) {111\k};
   }
   
    \foreach \k in {0,1,2}
         \node[leaf] (0110\k) at (-2*\n+\k*1*\n,\llb) {};
   
\end{scope}
\begin{scope}[>={Stealth[black]},
              every node/.style={fill=white,rectangle},
              every edge/.style={draw,thick}]
    \path [->] (0) edge node {0} (00);
    \path [->] (0) edge node {1} (01);
    \path [->] (00) edge node {$|\cN_0\r$} (000);    
    \path [->] (00) edge node {$\cN_1\r$} (001);    
     \path [->] (01) edge node {$|A\r$} (010);    
     \path [->] (01) edge node {$|B\r$} (011);    
     \path [->] (01) edge node {$|C\r$} (012);   
     
     \path [->] (010) edge node {0} (0100);    
     \path [->] (010) edge node {1} (0101);    
     \path [->] (011) edge node {0} (0110);    
     \path [->] (011) edge node {1} (0111);
    
    \foreach \k in {0,1,2}
     \path [->] (0100) edge node {$|$\k$\r$} (0110\k);    
     
     \path [->] (0111) edge node {$1$} (01110);    
     \path [->] (0111) edge node {$e^{i\phi}$} (01111);    
    \path [->] (0111) edge node {$e^{-i\phi}$} (01112);    
     
\end{scope}
\begin{scope}[every path/.style={thick,dashed}]
    \node (113) at (7*\n,\lop/2) {};
    \path  (01) edge node {} (113);    

    \foreach \l in {0,...,2}
    {
         \node (0101\l) at (0.5*\n+\l*0.5*\n,\llb-0.5*\m) {};
             \path  (0101) edge node {} (0101\l);   
         
     \node (0110\l) at (2*\n+\l*0.5*\n,\llb-0.5*\m) {};
             \path  (0110) edge node {} (0110\l);   
     }
     
    \node (01113) at (1*\n+3*2*\n,\le-0.5*\m) {};
    \path (0111) edge node {} (01113);    
\end{scope}
\end{scope}
\end{tikzpicture}
\caption{\label{fig:indexTree}
Index hierarchy for haCC (abbreviated upper part). Axis names
are indicated in yellow. Nodes on the respective levels are labeled by $J_l$, factor basis functions connect a node to 
the next-lower level.}
\end{figure}

In tRecX, any expansion is represented as a tree structure, starting from the total wave-function $\Psi$ at its root and the branches 
attached to each node of the tree are labeled by all indices down to the node's position in the tree, see Fig.~\ref{fig:indexTree}. 
The expansion coefficients
$c_\cN$, $c^A_i$, $c^A_\al$ are the leafs in a tree with separate branches for neutrals and channels and further in each channel for molecular orbitals $i$ and the numerical functions 
$\al$. Within each $\al$, magnetic quantum numbers $m$, angular momentum $l$,  the intervals $[r_n,r_{n+1}]$, and finally the individual Lagrange polynomials are themselves arranged in a tree. The tree-structure is recursive and induces, in a natural way, recursive algorithms for computing all pieces of the matrix. In the recursive algorithms one only needs the indices of the
branches at given node, but one never deals with the multi-indices that would describe the exact position in the tree.  
The discretization tree of figure \ref{fig:indexTree} is defined as \lcode{tRecX} input in the form
\\
\setlength{\tabcolsep}{3pt}
\begin{tabular}{rrrrrrr}
Axis:  subset, & name,& functions,&  order,& upperEnd,& nElem\\
        Neut,&Neutral,& CI[Neutral],&   2\\
        Chan,&   Ion,&     CI[Ion],&    3\\
    @Molecular,& Orbital,& CHEMICAL\\
@Perpendicular,& Phi,&     expIm,&      3\\
            &   Eta,&assocLegendre\{Phi\},&5\\
            &   Rn,&    polynomial,&    10,& 20,& 4\\
            &   Rn,&    polExp[0.5],&  25\\
\end{tabular}\\
\noindent
\revised{
which includes the radial axis with}
{
The first input line after \lcode{Axis} specifies that two neutral states should be used,
the second line specifies 
three ionic channels. These are the states, $|\cN_1\r$, $|\cN_2\r$ and $|A\r,|B\r,|C\r$, respectively, in Fig.~\ref{fig:indexTree}. 
The following lines define the expansion for the free electron at each ionic channel.
It consists of an \lcode{Orbital} branch, which contains all $I$ molecular orbitals and the $|\al\r$ basis with three functions $e^{im\phi}, m=0,1,-1$ on the 
\lcode{Phi} axis followed by \lcode{associatedLegendre} functions $P_l^{|m|}(\eta)$ with $l<5$. The axis name \lcode{Eta} refers to $\eta=\cos\th$.
The two lines with axis name \lcode{Rn} define the radial finite element functions
$\xi_n^i(r)$ (cf. Eq.~\ref{eq:alfaBasis}) with
4 equal size elements on $r\in[0,20]$ and 10 Lagrange polynomials on each element and 25
polynomials damped by $\exp(-0.5 r)$ on $[20,\infty)$. The \lcode{Rn} hierarchy level is not shown in Fig.~\ref{fig:indexTree}.
}

For realistic systems the efficiency of calculations profits from 
a detailed adjustment of radial and angular expansion near the atoms and fine-tuning using input as above can become demanding.
While this can still be done by the user, custom discretizations
for atoms and molecules are provided together with the corresponding chemical data and only global parameters like polynomial degree and minimal angular momentum need to be specified in a summary command \lcode{AxisSystem}.

\section{Classes specific for haCC}
\label{sec:operatorHaCC}
The most important classes in tRecX are discussed in \cite{scrinzi22-trecx}. The main operator class is  \lcode{OperatorTree}, whose construction, data, and application is parallelized. At the leafs of an \lcode{OperatorTree} one finds \lcode{class OperatorFloor}, which usually are moderate size linear operators that implement structure like diagonality or tensor-product form. In haCC, because of the exchange terms, all floors of the field-free Hamiltonian $\mH_0$ are full matrices. 

The complete Hamiltonian matrix $\mH$ appears as an \lcode{OperatorTree} at the root level that contains the $\mH_0$ and all time-dependent terms
of the mixed-gauge dipole interaction as its first level of branches. Each of these splits into the $2\times2$ blocks corresponding to the \lcode{Neutral} and 
\lcode{Ion} branches as in Fig.~\ref{fig:indexTree}. The scheme recursively continues on each of these 4 blocks until the floor levels of left and right indices are reached.

\subsection{\lcode{OperatorHaCC}}
This class implements the construction and application of operators as in Eq.~(\ref{eq:operatorHaCCblock}). It is derived from \lcode{OperatorTree}, but its \lcode{apply}-function is re-implemented as the application of, in sequence, $\mU\adj$,  $\mm$, and $\mU$. These three operators are each implemented as an ordinary \lcode{OperatorTree}. All terms of $\mm$ operate on an extended index space, where an auxiliary orbital node is attached to each 
\lcode{Orbital&Phi} branch in addition to \lcode{Orbital} and \lcode{Phi}, see Fig.~\ref{fig:indexTree}, and $\mU$ maps from this extended space to the original. The cost for that transformation is low and it is done only once for all terms in the operator, as
\beq
\mH(t) = \mU \left[\mh_0 + \sum_{k} f_k(t) \mh_k\right] \mU\adj,
\eeq
where $f_k(t)$ denote the time-dependent factors $\Ef_z(t)$, $A_z(t)$, etc.  

\subsection{\lcode{BasisOrbitalNumericalGaussian}}
Basis functions in tRecX are all derived from a base class \lcode{BasisAbstract} which requires the basis to have a \lcode{size()}. Derived classes are, e.g.
\lcode{BasisGrid} for a set of discrete points on the given axis, or \lcode{BasisExpIm} for $e^{im\varphi}$. Multi-dimensional basis functions can be defined by a 
set of numerical functions, which then are expressed on a discretization. Ref.~\cite{scrinzi22-trecx} contains a more comprehensive discussion. 

The class \lcode{BasisOrbitalNumericalGaussian} represents the molecular orbitals by an expansion into spherical harmonics and also by the original expansion into Gaussians. The spherical expansion uses the same radial element boundaries $r_n$ as the $\al$-basis Eq.~(\ref{eq:alfaBasis}), but the expansion size on each element is automatically increased until one obtains a converged basis $|\ga\r$, where overlap and kinetic energy agree with the exact Gaussian value to within a given tolerance. In that way, the tRecX standard scheme for operator definitions can be used for evaluating the matrix elements of any single-particle operators $S$ as 
\beq
\l i | S |\al\r=\sum_{\ga\ga'}\l i|\ga\r(\mo\inv1)_{\ga\ga'}\l\ga'|S|\al\r,
\eeq
where $\mo_{\ga\ga'}=\l \ga|\ga'\r$ is the FE-DVR  overlap for the $\ga$-basis.
For evaluating $\l\ga'|S|\al\r$ the standard tRecX scheme is used, where operators are defined in quasi-mathematical notation, 
see Ref.~\cite{scrinzi22-trecx}.
This enhances code flexibility and reduces programming errors.

\subsection{\lcode{InverseHaCC} and \lcode{InversePerp}}
The overlap consists of two independent blocks on $\cN\oplus\cI$ and $\cA$, respectively.
Pseudo-inverses for both blocks by the Woodbury formula are set up in \lcode{InverseHaCC}. 
On the top level, \lcode{InverseHybrid} implements Eq.~(\ref{eq:pseudoInverse}) for
the $\cN\oplus\cI$ block. The class  \lcode{InversePerp} implements $\cA\cA$  pseudo-inverse block  for the $|\alp\r$. 

\section{The tSurff and iSurff methods}
The method of time-dependent surface flux (tSurff) \cite{Tao2012,scrinzi12:tsurff} allows the efficient computation of photoelectron spectra using simulation volumes that are much smaller than the size to which the wave function expands during the interaction with the laser pulse. In tSurff one integrates the flux that leaves the simulation volume with appropriate time-dependent phases that account for the fact that electron momenta get modified by the laser field long after electrons have left the simulation box. The spectral amplitude accumulated from time $t=0$ until $t=T$ has the general form
\beq\label{eq:tsurff}
b(\vk,T)=\int_{0}^T \l \chi_\vk(t)|S(t)|\Psi(t)\r dt
\eeq
where $\Psi(t)$ is the time-dependent wave function, $S(t)$ is a time-dependent flux operator at radius $r=R_c$ with a modification due to the external field, and $\chi_\vk(t)$ are plane waves with time- and field-dependent phases. 
As $S(t)$ is non-zero only at $r=R_c$, all that is needed for computing spectra is the time-dependent flux at the boundary of the simulation volume $R_c$ until all relevant amplitude has left the simulation volume. A comprehensive discussion of tSurff for single-  and multi-electron emission can be found in Refs.~\cite{Tao2012,scrinzi12:tsurff}.

Together tSurff and infinite range exterior scaling (irECS) \cite{Scrinzi2010}, which is a mathematically rigorous and computationally efficient method for reflectionless absorption,  form the name-giving constituent methods of the tRecX=tSurff+irECS code.

A method for computing differential photoelectron spectra from the wave function at the end of the laser pulse had been proposed in Ref.~\cite{palacios07:spectra} and was used for the photo-emission spectrum from the $H_2$. That method relies on the solution of the resolvent equation for the full Hamiltonian.
An analogous method can be derived from Eq.~(\ref{eq:tsurff}): one observes that, without the external field, $\chi_\vk(t)=\exp[-i(t-T)|\vk|^2/2]\chi_\vk(T)$ corresponds to free motion, $S$ becomes time-independent and coincides with the standard flux operator, and $\Psi(t)=\exp[-i(t-T)H_0]\Psi(T)$ can be obtained by exponentiation. With this, the integral Eq.~(\ref{eq:tsurff}) from time $t=T$ to $t=\infty$ reduces to 
\beq\label{eq:isurff}
b(\vk)=\lim_{\ep\sd0}\l \chi_\vk(T)|S\Big(H_0-\frac{\vk^2}{2}+i\ep\Big)\inv1|\Psi(T)\r.
\eeq
Upon discretization $H_0\to \mH_0$ the limit may become ill-defined, which shows as numerical instability when $\vk^2/2$ approaches any  of the discrete
eigenvalues $E_i$ of $\mH_0$. However, with complex scaling the continuous spectrum of $H_0$ rotates into the lower complex plane and so do the 
eigenvalues $E_i$ of the complex scaled matrix $\mH_0$ and the inverse can be directly evaluated by setting  $\ep=0$.

The computational challenge for using this formula is that, for each photoelectron energy $E=\vk^2/2$ one needs to solve the linear system
\beq\label{eq:linsolve}
(\mH_0-E\mO)|\varphi\r = \mO|\Psi(T)\r
\eeq
where $\mO$ is the overlap matrix. For general $N\times N$ matrices $\mH_0$, solving Eq.~(\ref{eq:linsolve}) scales as $N^3$.
When $\mH_0$ is block diagonal, as in the case of atoms, the effort reduces accordingly. In tRecX, rather than repeatedly solving the linear system, we use a spectral decomposition of the complex scaled $\mH_0$, which is a one-time $\order{N^3}$ calculation. The decomposition has the form
\beq\label{eq:specRep}
\exp\left(-it\mO_\ep\inv1\mH_0\right) = \mR\exp\left(-i t\md\right) \mL^T\mO,
\eeq
where $\mR$ and $\mL$ are  the matrices of right and left eigenvectors of generalized eigenvalue equations for 
the Hamiltonian matrix $\mH_0$ 
\beq
\mH_0 \mR = \mO \mR \md,\qquad  \mL^T\mH_0 =  \md\mL^T \mO
\eeq
with the overlap matrix $\mO$, its pseudo-inverse $\mO_\ep\inv1$, Eq.~(\ref{eq:pseudoInverse}), and the diagonal matrix of eigenvalues $\md$. 
The distinction of right and left vectors is needed, as with absorption $\mH_0$ is not hermitian $\mH_0\neq\mH_0\adj$. 
For some bases, one may maintain complex symmetry $\mH_0=\mH_0^T$, in which case $\mL=\mR$. Left and right eigenvectors are orthogonal \wrt $\mO$ in the sense
\beq
\mL^T\mO \mR  = \mP_\ep,
\eeq
where $\mP_\ep$ is the projector onto the non-singular subspace, Eg.~(\ref{eq:projEps}).


The combination of tSurff with the time-independent integral Eq.~(\ref{eq:isurff}) was first proposed in Ref.~\cite{morales16:isurf} 
under the name of iSurff. It is particularly useful for the 
low-energy part of the spectrum, for which wave-function density leaves the simulation box slowly and tSurff needs to be 
propagated for a long time after the end of the pulse. Another application is the presence of long-lived resonances, which may exhibit 
time-scales that far exceed the pulse duration. When comparing iSurff to  Ref.~\cite{palacios07:spectra} one may use much smaller simulation boxes. This reduces both, the propagation time through the pulse and the time needed for solving the system spectral decomposition Eq.~(\ref{eq:specRep}) or, alternatively, the linear system solving Eq.~(\ref{eq:linsolve}).

The present release of tRecX includes \lcode{iSurff} as a standard routine which is invoked by the input  \lcode{Spectrum:iSurff=true}.
The implementation uses classes from tSurff for computing surface values and the 
momentum-dependent phases, and maps to the momentum grid. For the spectral decomposition, the class \lcode{DiscretizationSpectral}
is employed.

\section{Parallelization and scaling}
\label{sec:parallel}

tRecX is parallelized using MPI but it will also compile without MPI, if no MPI implementation is detected on the system. Finest grains for parallelization
are sections of the coefficient vectors of length $10\sim 100$, the coefficient ``floors'', and the corresponding \lcode{OperatorFloor} blocks.
Each coefficient floor is hosted on one process, and the  \lcode{OperatorFloor} is ``owned'' by the process that hosts either its input or its output 
coefficients. The distribution of \lcode{OperatorFloor} is otherwise arbitrary and they are moved in order to achieve best load balancing. 
As the application cost of \lcode{OperatorFloor} varies largely due to varying sizes and partially sparse structure, load balancing is achieved
by redistributing the \lcode{OperatorFloor}'s after setup, based on timing the application cost for each floor. 

When using haCC, where $\mH_0$ is full because of exchange and dominates application time, column-wise distribution gives good load balancing, if one adjusts column boundaries using the actual timings. The time-dependent parts have much lower application cost and use the same parallel layout as $\mH_0$.  

\subsection{Class \lcode{FlattenedOperator}} \revised{}{This class controls distribution and parallel application of an \lcode{OperatorTree}. It ``flattens'' the
original tree structure into a vector of \lcode{OperatorFloor}'s, which allows equal treatement of all \lcode{OperatorFloor}'s, independent of their position in the 
\lcode{OperatorTree}. The mapping is by pointers, which allows keeping the original tree structure for reference on all processes 
without creating copies of the \lcode{OperatorFloor}'s.  For each \lcode{OperatorFloor}, only its owner process has actual data, while there are 
dummies with negligible memory on all others.}
The more complex classes \lcode{OperatorHaCC} or \lcode{InverseHybrid} are parallelized by flattening all their \lcode{OperatorTree} factors.

\subsection{Operator setup} 
Considerable compute resources are consumed by 
the computation of matrix elements of two-particle operators on $\cA$ and the mixed Gaussian-numerical matrix elements between $\cA$ and $\cI$, where  up to  three-particle reduced reaction density matrices enter (see Sec.~\ref{sec:computation}).  As to be expected, the most time-consuming part is the computation of exchange on the $\cA$-subspace.
 The integrals are calculated by representing Coulomb repulsion of electrons $|\vr-\vr'|\inv1$ in a multipole expansion for the angular degrees of freedom and a quadrature grid in the two radial coordinates $(r,r')$ that is exact for the basis, see Ref.~\cite{zielinski16:doubleionization} for details. Parallelization is by distributing the pairs of radial quadrature points on a given FE-DVR $mn$-patch $[r_m,r_{m+1}]\times[r'_n,r'_{n+1}]$ across the parallel 
processes. This strategy was chosen over distributing whole patches, as the number of angular momenta and quadrature points can vary greatly between different $mn$-patches and load balancing is easier to achieve between the points within the same $mn$-patch. Variations of that strategy were used for parallelizing the evaluation of all other two-electron integrals. As setup still can take significant time, one can save the operators for a
given expansion and re-use it for re-runs within the same expansion, where hashing ensures access to the correct matrices. 

\subsection{Scaling}
Floating operations and communication are dominated by $\mH_0$, which is a full matrix because of exchange, whereas the time-dependent operators are implemented by sparse and partially local operations, as shown in Sec.~\ref{sec:operatorImplementation}. At any parallel run, tRecX will print measured load-balancing and the communication patterns of the major operators, which helps to detect load balancing problems. With the full matrix $\mH_0$, load-balancing in haCC applications is typically within a few percent, with slightly poorer balancing for the remaining parts of the operator, which, however, consume only a fraction of the compute time. 

\begin{figure}
\includegraphics[width=0.7\textwidth]{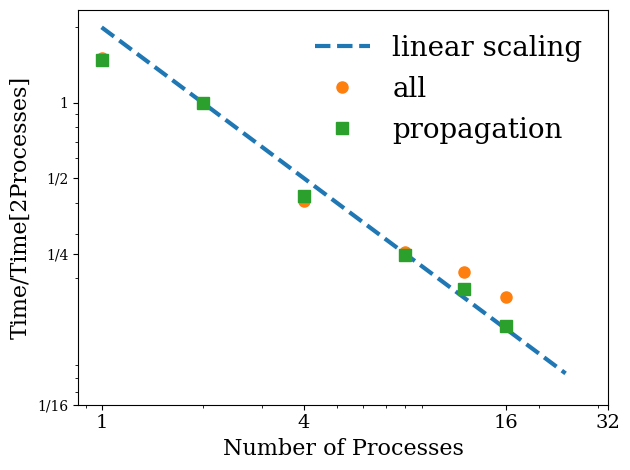}
\caption{\label{fig:scaling} Scaling of total computation and setup times for medium size haCC calculation. Results are comparable to the scaling reported in Ref.~\cite{scrinzi22-trecx}. \revised{}{Times are normalzed to the 2-process calculation, as some algorithms differ in the strictly scalar code with a single process.}
}
\end{figure}

Application of the full matrix and the limit to granularity in terms of the \lcode{OperatorFloor}'s allows favorable scaling only up to about
16 processes, depending on the problem size and hardware. Fig.~\ref{fig:scaling} shows the scaling at fixed problem size (``strong scaling'') for the example of photo-ionization of the $CO$ molecule.

\section{Input, documentation, output, and compilation} 

Input is defined through files and handled internally by \lcode{class ReadInput}, which also supports documentation of the code, as discussed in some detail in Ref.~\cite{scrinzi22-trecx}. Otherwise it is conscious policy {\em not} to provide a separate document that describes the code, other than the present article and Ref.~\cite{scrinzi22-trecx} that outline the capabilities of the code and its structure. Code use is systematically introduced by the tutorials that are provided with each release. The most important of these
are discussed below and in Ref.~\cite{scrinzi22-trecx}. Documentation of the input is written into the code using the class \lcode{ReadInput}. A short reference for the input options can be displayed by running tRecX without any arguments, more extended help for each input category is shown by the \lcode{-h} flag, e.g. \lcode{./tRecX -h=Axis} for a brief summary for every entry related to the input category \lcode{Axis}. Finally, for many inputs longer explanatory texts are written right with the respective \lcode{ReadInput:read} command. These are routinely processed during every code sanity check and automatically generate an up-to-date \lcode{UserManual.pdf} that is published with the code.
For more technical purposes the \lcode{C++} is marked by Doxygen \cite{doxygen} documentation and further in-line comments.

\paragraph{Run directory} Each time the code is started with a given input file name \lcode{SomeName.inp} it generates a ``run directory'' \lcode{SomeName/xxxx}, where the 4-digit code \lcode{xxxx} is the first available starting from \lcode{0000}. A copy of the original input and all outputs are placed into that directory.

A few python scripts for summary display of multiple runs, plotting and comparing spectra and submitting multiple runs to a queuing system are provided with the package. A brief description is given in \lcode{UserManual}, up-to-date info is displayed by running the scripts without arguments.  

\revised{}{
Compilation of the code is by \lcode{cmake/make}, which has been tested on both, Linux-based and Mac systems, but only the Linux version is actively maintained. The code makes use of a range of public domain libraries. In order to secure matching library versions the \lcode{Eigen} template library \cite{eigen} and \lcode{ALGLIB} \cite{alglib} are supplied with the distribution. Up-to-date build instructions and protocols of successful builds are included with the \lcode{README.md} file. 
}

\section{Usage examples}
\label{sec:applications}

Up-to-date information about possible applications in the form of tutorials is shipped with the code  \cite{scrinzi22-trecx} and is available also in the AMOS portal \cite{AMOSgateway}.
For the discussion of \revised{the most exemplary}{} applications already available in the first release we refer the reader to \cite{scrinzi22-trecx}. The following applications all involve haCC. 

\revised{}{
Computation times strongly depend on the problem. In particular, long wave-lengths require higher angular momenta and longer propagation times. In general, because of $\mH_0$ being a full matrix, times grow quadratically with discretization size. 
The examples below were chosen for easy reproducibility with compute times varying between $\sim 5$~minutes at wave-length of 20 nm  on 4 cores of a laptop 
for the largest calculation shown below in Fig~\ref{fig:HeXUV} to  three hours at 400 nm on 8 cores of a small cluster for the largest calculation of Fig.~\ref{fig:CO400}. 
}

\subsection{Photo-emission and high harmonic generation with Helium}
\label{sec:exampleHe}

While the Helium atom has significant correlation in its ground state, strong field processes are assumed to be effectively single-electron in many cases. We examine this assumption for three different situations: photoelectron spectra generated by a short pulse at the extreme ultraviolet (XUV) wave length,  strong-field ionization at visible wave-length and high harmonic generation.

A minimal input for computing photo-emission is (see \lcode{tutorial/200HeHaCC}) 
\begin{lstlisting}
Chemical: data=../Chemical/He19_Standard
AxisSystem: lMax=4
Spectrum: radialPoints=200, plot=channels, iSurff=true
Laser: shape, I(W/cm2), FWHM,   lambda(nm)
       cos8,   1.e16, 3. OptCyc, 20
\end{lstlisting}
The input \lcode{AxisSystem} directs the code to use a default discretization that is specified together with the pre-processed quantum chemical data and that is safe for standard applications. As with any such calculation, convergence needs to be confirmed by varying the discretization. The maximal angular momentum \lcode{AxisSystem:lMax} must always be specified. For the given laser parameters, \lcode{lMax=4} is sufficient, as hardly more than 1 photon is absorbed.  The spectrum will be computed on a momentum grid of 200 \lcode{radialPoints}, the plot will be for the channel-wise energy spectrum, and \lcode{iSurff} will be used. Laser pulse parameters are intensity $10^{16}W/cm^2$, full width at half maximum (FWHM) of 3 optical cycles and central wave length of 20 nm. \revised{Unless indicated in the name of the input item,}{Inputs are understood to be in atomic units (au). }
\revised{}{An exception are names  that explicitly indicate the input units, like \lcode{I(W/cm2)} and \lcode{lambda(nm)}. However, one can specify the input unit next to the input value.} In this example, we have chosen to specify  \lcode{FWHM} in optical cycles \lcode{OptCyc} rather than the default au. \revised{}{Conversily one could, e.g., give the intensity as \lcode{1 au}, which would correspond to $\approx3.51\,10^{16}W/cm^2$. } Many other input units like seconds, meters, etc. can be given in a similar way, see Ref.~\cite{scrinzi22-trecx} for a more comprehensive discussion.

\revised
{
}
{
The details of the discretization generated by the \lcode{AxisSystem} directive are written to a file \lcode{axisSystem} in the run directory. That file has the format of a standard tRecX input. Most importantly, it contains an \lcode{Axis} definition as discussed in Sec.~\ref{sec:representation} where the radial sections for the $\al$-basis are chosen to match the given \lcode{Chemical:data}. Some of the parameters of that discretization can be changed by specifying, e.g. \lcode{AxisSystem:nChan} for the number of channels or \lcode{AxisSystem:rMax} for the radius of the simulation volume. More experienced users can include the file \lcode{axisSystem} into the input and change the given values, e.g., when performing a more detailed convergence study. 
}

By default, the calculation is with a single channel. The number of channels is set by
\begin{lstlisting}
AxisSystem: nChan=2
\end{lstlisting}
for the ground  and the  first excited ionic states. Matrix sizes grow with the square of the number of channels, i.e. the 2-channel calculation will take about 4 times the resources of the single channel calculation. For \lcode{iSurff} 
the growth is with the 3rd power, as the eigenproblem for the full matrix needs to be solved. 

\begin{figure}[h]
\includegraphics[width=0.6\textwidth]{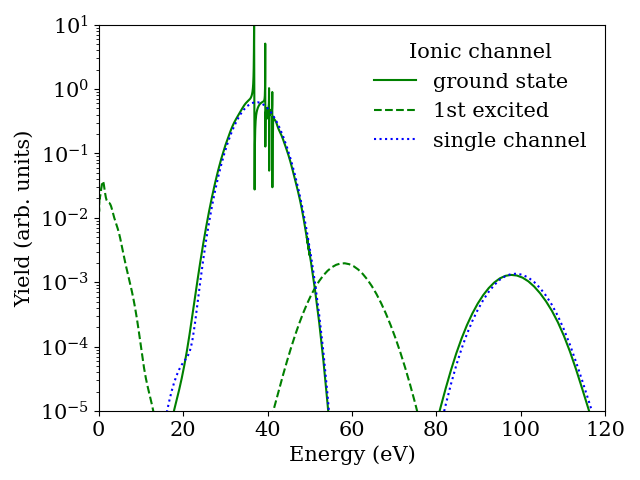}
\caption{\label{fig:HeXUV} Photo-emission from the He atom generated by a short intense pulse at a central wave length of $20\nm$, intensity of $10^{16}W/cm^2$, FWHM of 3 optical cycles and a $\cos^8$ pulse envelope.}
\end{figure}

Fig.~\ref{fig:HeXUV} shows the photo-emission spectra for a single- and a two-channel calculation. 
The broad features are the single-photon peaks for the ground and first excited ionization channel. On top of the first peak of 
the ground state channel there is a group of 2s-np resonances due to the coupling to the excited channel, see Sec.~\ref{sec:isurff}. Except for that, at the given parameters, the only multi-channel effects in the spectrum is a small shift to lower energies in the two-channel calculation, due to a slightly lower ground state energy in the multi-channel calculation.

When doing several calculations with the same discretization one can add the lines 
\begin{lstlisting}
Cache:
hamiltonian[He19] interaction
\end{lstlisting}
which specifies that operators will be cached in a directory defaulting to \lcode{CACHE_TRECX} with filenames beginning with 
\lcode{He19} and unique hash indicating the actual discretization.

At longer wave length, one needs to increase the size of the angular momentum expansion. We show calculations at 400 nm wave length, 
for which we adjust the input lines to (see \lcode{tutorial/201He400})
\begin{lstlisting}
Dipole: velocity, length
AxisSystem: lMax=25
Laser: shape, I(W/cm2),   FWHM,   lambda(nm)
         cos2, 2.e14,  6. OptCyc, 400
\end{lstlisting}
The wavelength of 400 nm is in the visible, pulse duration now 6 optical cycles in order to get well-separated photo-emission peaks. Angular momenta have been increased to 25, as this is in the truly multi-photon regime.
Also, we here use a $\cos^2$ pulse envelope. When used with very short pulses, the $\cos^2$ envelope can create artefacts, about which tRecX will issue a warning. However, the pulse shape is popular, and the artefacts are unimportant for the present demonstration purpose. 
The additional line \lcode{Dipole:velocity,length} will cause the dipole expectation values to be computed in, both, length- and velocity-form, from which harmonic responses are computed automatically. The acceleration form 
of the dipole cannot be used with haCC, because the code cannot form the derivative of the mean-field and exchange potentials.

\begin{figure}
\includegraphics[width=0.48\textwidth]{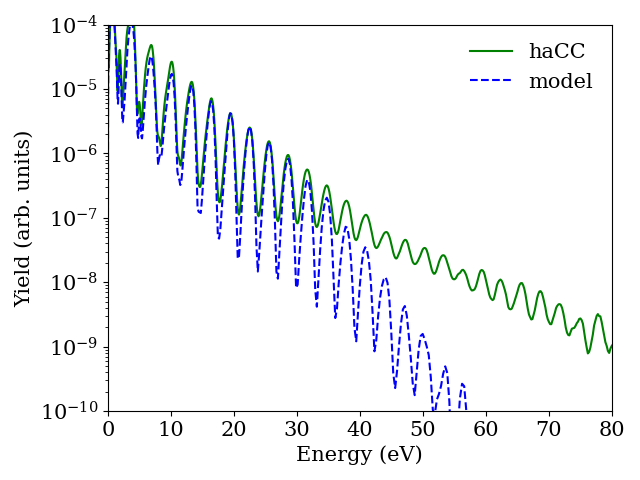}
\includegraphics[width=0.48\textwidth]{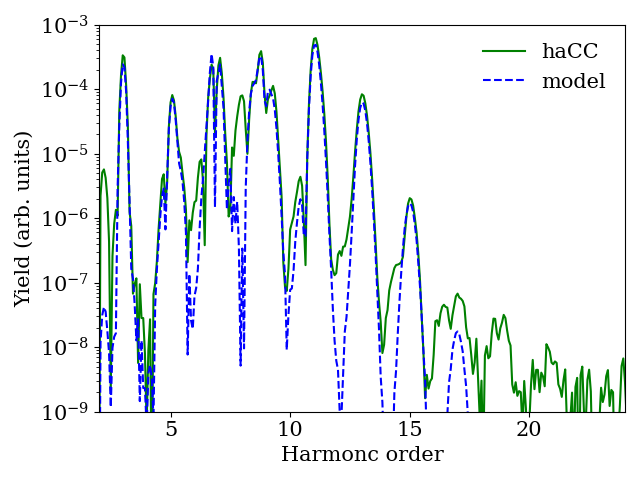}
\caption{\label{fig:He400} photoelectron (left panel) and high harmonic spectra (right panel) emitted from Helium in laser pulse with wave length 400 nm, peak intensity $2\times 10^{14}W/cm^2$ and 6 optical cycles pulse duration. Single-electron model (blue) vs. haCC (green).
}
\end{figure}
Figs.~\ref{fig:He400}(a) and (b) show photo-emission harmonic spectra for this case. The haCC calculations are numerically converged to within $\lesssim10\%$ relative accuracy for the peak maxima in the range shown. For comparison, the figure includes a single-electron calculation with the model potential
\beq
V(r)=-\frac{1+\exp(-2.1405 r)}{r},
\eeq
where the screening constant  is adjusted for an ionization potential of $0.90186 \,a.u.$ which matches the ionization potential obtained out of the COLUMBUS calculations. Important differences appear in the high-energy part of photoelectron emission and 
in the noise level of the harmonic spectra. The haCC calculation shows significantly more yield at high electron energies. Without further interpreting the result at this point, we observe that the deviations become dramatic beyond photoelectron energies of $30\,eV$, which is the well-known 10~$U_p$ cutoff for photo-emission by single recollision \cite{paulus94:10up}. The haCC calculation includes correlation and anti-symmetrization, but the given difference may also reflect the different behavior of the electron wave functions near the nucleus in the two calculations.

\subsection{Convergence and accuracies: photo-emission from $CO$}
\label{sec:convergence}

\begin{figure}[h]
\includegraphics[width=0.7\textwidth]{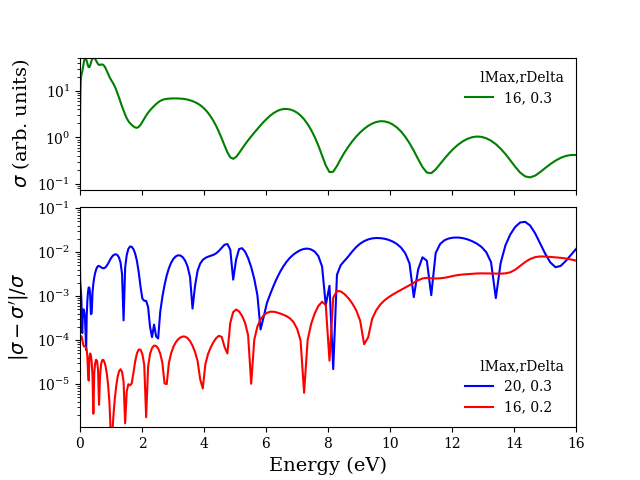}
\caption{\label{fig:CO400} Convergence of the photoelectron spectrum for $CO$ in a 6-cycle laser pulse at wave length 400 nm and peak intensity $10^{14}W/cm^2$. Spectrum $\si$ (upper panel) and relative difference $|\si-\si'|/\si$ (lower panel) 
comparing $\si$ with \lcode{lMax=20} angular momenta and the default radial spacing of \lcode{rDelta=0.3} with spectra $\si'$ with 16 angular momenta (blue) and radial spacing 0.2 (red), respectively.}
\end{figure}
\label{sec:CO}
For the assessment of the accuracy of a tRecX calculation one needs to examine the numerical convergence of a given observable \wrt 
discretization parameters. We discuss this here on the example of the photoelectron spectrum of the diatomic $CO$ in a laser pulse of 
400 nm wavelength, 3 optical cycles FWHM pulse duration and intensity of $10^{14}W/cm^2$. With the inputs
\begin{lstlisting}
Chemical: data=../Chemical/COlarge5_Standard
AxisSystem: lMax=16
Spectrum: radialPoints=200, plot=channels
Laser: shape, I(W/cm2), FWHM,  lambda(nm)
        cos2,   1.e14, 3. OptCyc, 400
TimePropagation: end,     print
              4 OptCyc, 1/8 OptCyc
\end{lstlisting}
we obtain the photo-emission spectra from $CO$ as shown in Fig.~\ref{fig:CO400}. 

\paragraph{Convergence \wrt angular momenta and radial discretization} Figure~\ref{fig:CO400} shows the relative differences of the spectrum compared to  two calculations with more angular momenta \lcode{lMax=20} and higher density of DVR points on the radial axis, which is set by \lcode{AxisSystem:rDelta=0.2} compared to the default value  of 0.3. Based on these relative differences we estimate that the calculation is numerically converged \wrt to these parameters to about 1\% relative accuracy at the respective photoelectron peak maxima. 

\begin{figure}[h]
\includegraphics[width=0.7\textwidth]{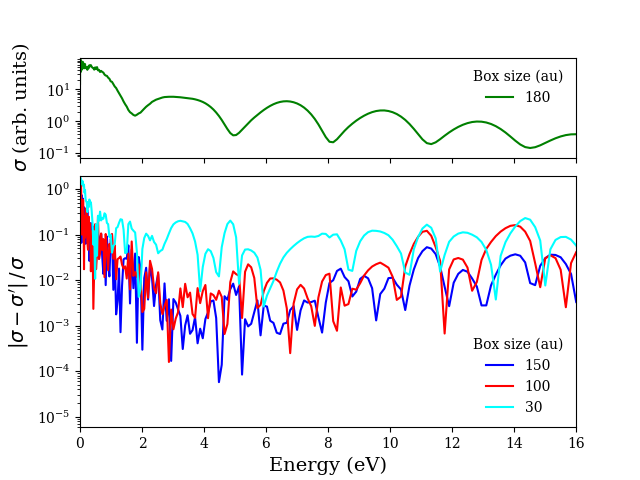}
\caption{\label{fig:CObox} Long-range Coulomb effects on photoelectrons. Comparing a calculation with box size 180 to calculations with sizes from 30 to 150 au. Up to box size of $\sim 50$ errors are on the scale of 10\%. By the convergence from box sizes of 180 through 150 to 100, one estimates an accuracy at photoelectron peak heights at box size 180 of $\lesssim1\%$, except near threshold.}
\end{figure}

\paragraph{Long-range Coulomb effects} The long-range Coulomb potential invariably affects photoelectron spectra. This can be exposed by increasing the radius of the simulation volume over the default value of 30~au (for $CO$) 
\begin{lstlisting}
AxisSystem: rMax=50
\end{lstlisting}
Note that there is a correlation between the radius and the number of required angular momenta: to some extent the growth of the 
volume is two-dimensional in case of cylindrical symmetry and one needs to re-establish angular convergence when using larger box size.
Fig.~\ref{fig:CObox} shows changes of about 10\% throughout the main part of the spectrum as one decreases the simulation box size from a maximum of 180 au to 30.  As to be expected, differences are largest at low energies that are most affected by the weak, but long-range Coulomb interaction. At energies $\gtrsim 2\,eV$ the calculation at box size 180 can be assumed to be converged to $\lesssim 1\%$ in relative accuracy. With the effect being due to the Coulomb potential, the box size needed for a given accuracy in a system with a given ionization potential can be estimated using a hydrogen-like model with the effective charge adjusted to match the ionization potential.  

Clearly, no general accuracy statements can be made across all laser parameters and observables. As in any such calculation, careful investigation of convergence for a given purpose remains the user's responsibility.  

The file \lcode{axisSystem} contains the definition of the discretization near the nuclei. The values provided have been tested to be safe for about 10\% accuracies for photoelectron spectra in the range from XUV to 800 nm wavelength and intensities that lead to only partial ionization. In the present calculation convergence \wrt to these further parameters is well below 1\%. When calculating at more extreme parameters, longer wave length or higher intensities, one also needs to test convergence \wrt the discretization near the nuclei. For that one would replace the input \lcode{AxisSystem} by the contents of the \lcode{axisSystem} file and examine sensitivity of the results to the various additional discretization parameters. A list of the relevant convergence parameters and explanations of their meaning can be found in the \lcode{UserManual}.

\subsection{Application of iSurff: resonances in $He$ and $N_2$}
\label{sec:isurff}

\begin{figure}[h]
\includegraphics[width=0.49\textwidth]{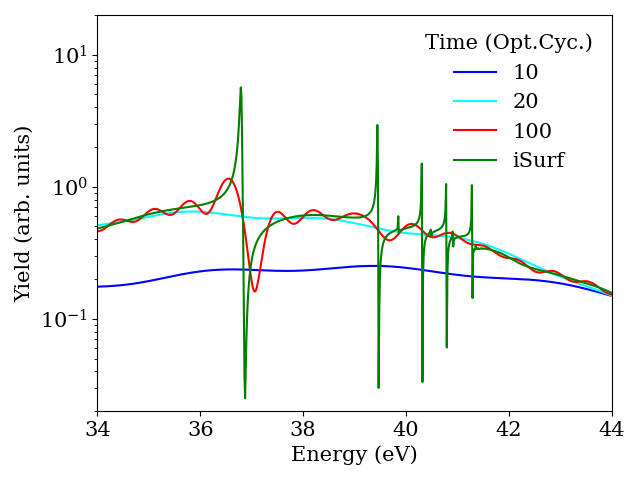}
\includegraphics[width=0.49\textwidth]{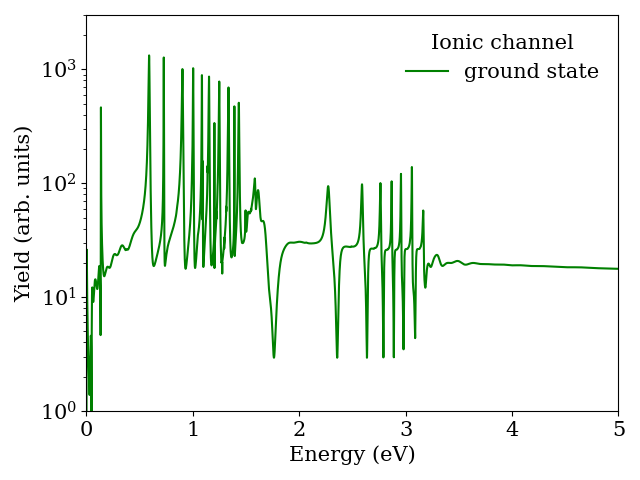}
\caption{\label{fig:resonances} Applications of iSurff: resonances due to doubly excited states. Left panel: detail of Fig.~\ref{fig:HeXUV} obtained with iSurff  (green), comparing to pure tSurff with finite propagation times. Right panel: resonances in the photo-ionization of $N_2$ into the ground state channel by a single-cycle pulse at 68~nm nominal wave length. }
\end{figure}
Fig.~\ref{fig:resonances} shows resonances the ground state channel for the $He$ atom and the $N_2$ molecule. The resonances arise from doubly excited states in the excited ionic channels. Because of the long lifetime of the resonances, one needs to use \lcode{iSurff=true} for resolving the peaks, which amounts to integration to infinite times. 

In the He calculation (left panel) we include pure tSurff spectra, obtained with 
the inputs
\begin{lstlisting}
Spectrum: iSurff=false
TimePropagation: end=10 OptCyc
\end{lstlisting}
with longer propagation times up to \lcode{end=100} optical cycles. 
At \lcode{end=10} optical cycles, significant part of the electron probability has not passed the surface, with 20 cycles the broad peak is
essentially complete, but only at times $\sim100$ optical cycles the peaks from the slowly decaying resonances start to appear.
This illustrates how the long-lived resonances are typical applications for \lcode{iSurff}.

The right panel shows photo-emission from the $N_2$ molecule at wave-length of 68~nm and pulse duration of 1~optical cycles FWHM, with two groups of resonances which correspond to double excitations above the first and second ionization thresholds. The calculation is included here for showing the capabilities of tRecX and the further discussion of the resonances will be presented elsewhere.

\subsection{The tRecX GUI on the AMOS Gateway}

\begin{figure}
\includegraphics[width=0.99\textwidth]{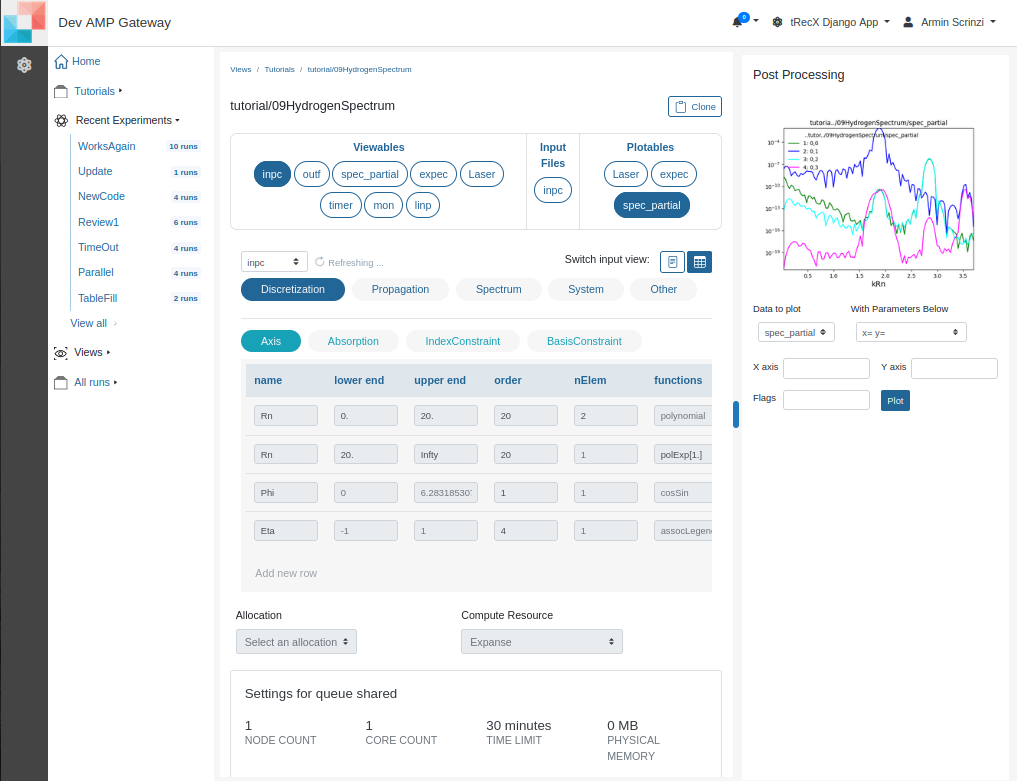}
\caption{\label{fig:amos} The GUI for tRecX on the AMOS Gateway. Input items sorted by category, input and output files, and graphical display of the results can be accessed. Basic help on the items appears upon hovering over the buttons. Submission to a selection of resources is controlled in the lower part of the screen. Runs are grouped into ``Experiments'' shown in the left panel.
}
\end{figure}

An HTML-based GUI was designed for tRecX and implemented in the AMOS Gateway \cite{AMOSgateway} that hosts several codes for atomic, molecular and optical sciences. The codes and a range of resources can be accessed through the gateway in a uniform way. The tRecX GUI is primarily for low-threshold access by first-time users. It provides all essential tutorials, that can be amended  and changed all the way to completely new calculations. Results can be downloaded and graphics can be produced by interactive access to the \lcode{plot.py} script provided with the tRecX release. Fig.~\ref{fig:amos} shows a screenshot of the interface. \revised{}{Access to a free acount to the AMOS Gateway, with limited resources, will be granted after review by one of administrators.}

\section{Conclusions}

The haCC implementation in the present major update of tRecX enables a wider range of specialist users to perform correlated multi-electron calculations for atoms and small molecules. The tRecX framework ensures a uniform and error-safe handling of the calculations. By their nature, such calculations require, in general, significant compute \revised{resources}{times}, except for the simplest atomic cases. 

The code uses a new variant of haCC that makes explicit use of the molecular orbitals of the underlying CI calculation. This enhances 
the multi-center capability of haCC compared to its original formulation in Ref.~\cite{majety15:hacc}. An outline of that re-formulated method was presented, 
which includes the general approach to computing matrix elements, control of near linear dependencies by Woodbury-type decompositions, methods for handling complexity of haCC by recursive schemes, and efficient application of the operators. 

We have demonstrated haCC on the example of laser-ionization of the Helium atom and diatomics. Extension to non-linear molecules is planed for a
future upgrade. The important question of convergence has been discussed using the non-trivial example of the $CO$ molecule by giving examples
for the most important convergence checks, with a more complete list in tRecX's \lcode{UserManual.pdf}.

A further major new feature of this release is the iSurff \cite{morales16:isurf} method. The method was re-derived within the existing tSurff formalism 
and its relation to a well-established earlier method \cite{palacios07:spectra} was pointed out.  The resolution of long-lived resonances in broad-band XUV excitation of Helium was given as an example.

The new release contains a range of gradual improvements, like the possibility of saving operators for re-use in later calculations, simplifications for in- and output, extension of the automated documentation, a revision of the parallelization strategy, and improved scripts for data presentation. 

The implementation philosophy of tRecX was maintained: it is designed for large scale applications with verifiable accuracies, as well as for training, education, and community development. A designated part of the tRecX development is to ensure user experience that is acceptable to a somewhat wider range of specialist users, including experimentalists who want to generate standard results or study simple models as well as theorist
with more complex demands. We consider error safe and intuitive input, extensive consistency checks, and structurally enforced documentation as essential for achieving that goal. 

\section*{Author contributions}
{\em H. Chundayil:} Methodology, Software, Validation, Writing - Review \& Editing; 
{\em V.P. Majety:} Conceptualization, Methodology, Software, Writing - Review \& Editing;
{\em A. Scrinzi:} Conceptualization, Methodology, Software, Writing - Original Draft, Visualization.

\section*{Acknowledgment}
The code was originally developed by the authors and Alejandro Zielinski, with 
further contributions by, in alphabetic order, Christoph Berger, Jonas Bucher, Florian Egli, Jacob Liss, Mattia Lupetti, David Z. Manrique,Jonathan Rohland,  J\"orn St\"ohler, Andreas Swoboda, Hakon Volkmann, Markus and Michael Weinmueller,  and Jinzhen Zhu. V.P.M. acknowleges support from the SERB startup research grant.

\bibliographystyle{unsrt}
\bibliography{../../bibliography/photonics_theory}

\end{document}